\documentclass{optica-article}

\journal{oe}

\begin{document}

\title{Low-loss polarization control in fiber systems for quantum computation}

\author{Tomohiro Nakamura,\authormark{1} Takefumi Nomura,\authormark{1} Mamoru Endo,\authormark{1,2} He Ruofan,\authormark{1} Takahiro Kashiwazaki,\authormark{3} Takeshi Umeki,\authormark{3} Jun-ichi Yoshikawa,\authormark{2} and Akira Furusawa\authormark{1,2,*}}

\address{\authormark{1}Department of Applied Physics, School of Engineering, The University of Tokyo, 7-3-1 Hongo, Bunkyo-ku, Tokyo 113-8656,
Japan\\
\authormark{2}Optical Quantum Computing Research Team, RIKEN Center for Quantum Computing, 2-1 Hirosawa, Wako, Saitama 351-0198, Japan\\
\authormark{3}NTT Device Technology Labs, NTT Corporation, 3-1 Morinosato Wakamiya, Atsugi, Kanagawa 243-0198, Japan}

\email{\authormark{*}akiraf@ap.t.u-tokyo.ac.jp} 



\begin{abstract}
Optical quantum information processing exploits interference of quantum light. However, when the interferometer is composed of optical fibers, degradation of interference visibility due to the finite polarization extinction ratio becomes a problem. Here we propose a method to optimize interference visibility by controlling the polarizations to a crosspoint of two circular trajectories on the Poincar\'{e} sphere. Our method maximizes visibility with low optical loss, which is essential for quantum light, by using fiber stretchers as polarization controllers. We also experimentally demonstrate our method, where the visibility was maintained basically above 99.9\% for three hours using fiber stretchers with an optical loss of 0.02~dB (0.5\%). Our method makes fiber systems promising for practical fault-tolerant optical quantum computers.
\end{abstract}


\section{Introduction} \label{sec1}
Quantum computers are gathering more attentions as next-generation technologies, and extensively researched with various physical systems. Optical implementation is one of the promising candidates. Optical quantum computer will be composed of beamsplitter networks by which quantum light interacts, followed by detectors. In particular, cluster states, which are entangled states that serve as resources for measurement-based quantum computation, are generated by beamsplitter interactions of squeezed states. In recent years, ultra-large-scale cluster states are experimentally demonstrated by employing the time-domain multiplexing method, which utilizes Mach-Zehnder interferometers with asymmetric lengths of arms\cite{Menicucci2011, Yokoyama2013, Yoshikawa2016, Asavanant2019, Larsen2019_2d}. The time-domain multiplexing method takes advantage of the flying feature of light. The demonstrated size of entanglement, therefore, is several orders of magnitude larger than that with other quantum systems such as superconducting qubits, showing the promise of optical quantum computing.

In order to exploit the strong quantumness of light, high interference visibilities are required in the interferometers. For this purpose, both the spatial mode and the polarization mode of the light beams should be matched. In the case of free-space systems, even though it is possible to match the spatial mode of all light beams at all interference points, it is time-consuming. Furthermore, the beam positions shift over time to reduce interference visibility, even though this problem can be solved by introducing auto-alignment systems. Due to time constraints, visibilities are often compromised, e.g., the experiment in Ref.~\cite{Yokoyama2013} was performed with 98\% visibilities. On the other hand, optical fibers are suitable for constructing maintenance-free systems. Using optical fibers, it is possible to achieve almost 100\% visibilities as long as the polarization in the fiber matches. 

However, even if polarization-maintaining fibers are used in the fiber systems, it is difficult to perfectly maintain polarizations due to the finite polarization extinction ratio (PER) of the fiber components. Every fiber component such as a fiber beamsplitter, a connector, or a connection point with fiber fusion shows a finite PER. When an interferometer is constructed using fibers, visibility, therefore, degrades due to polarization mismatch, which is called visibility fading\cite{tur1995, Huang2009}.

For classical light, visibilities can be improved by using commercial inline-type fiber polarization controllers. There are basically two types of fiber polarization controllers. In one type, a non-polarization-maintaining fiber is wound around spools and the angle of the spools are adjusted to cause appropriate stress-induced birefringence, resulting in desired polarization change\cite{Koehler1985}. In the other type, a non-polarization-maintaining fiber is stressed by piezo actuators to cause birefringence for desired change of polarization\cite{Johnson1979}. In these polarization controllers, non-polarization-maintaining fibers are subjected to stresses, which generate microbends to cause optical losses [typically 0.08~\text{dB} (1.8\%) or higher]\cite{Jay2010}. However, for classical light, these levels of optical losses are not serious problems.

For quantum light, polarization controllers are also needed to solve the visibility fading problem. However, since the quantum light is vulnerable to optical losses, the polarization controllers must have low optical losses. For example, there is a study showing that $-10$~\text{dB} of squeezing is required for fault-tolerant quantum computation\cite{Fukui2018}. In order to achieve this, the losses of the entire systems must be sufficiently lower than 0.5~\text{dB} (10\%).

\begin{figure}[tb]
\centering
\includegraphics[width=\linewidth]{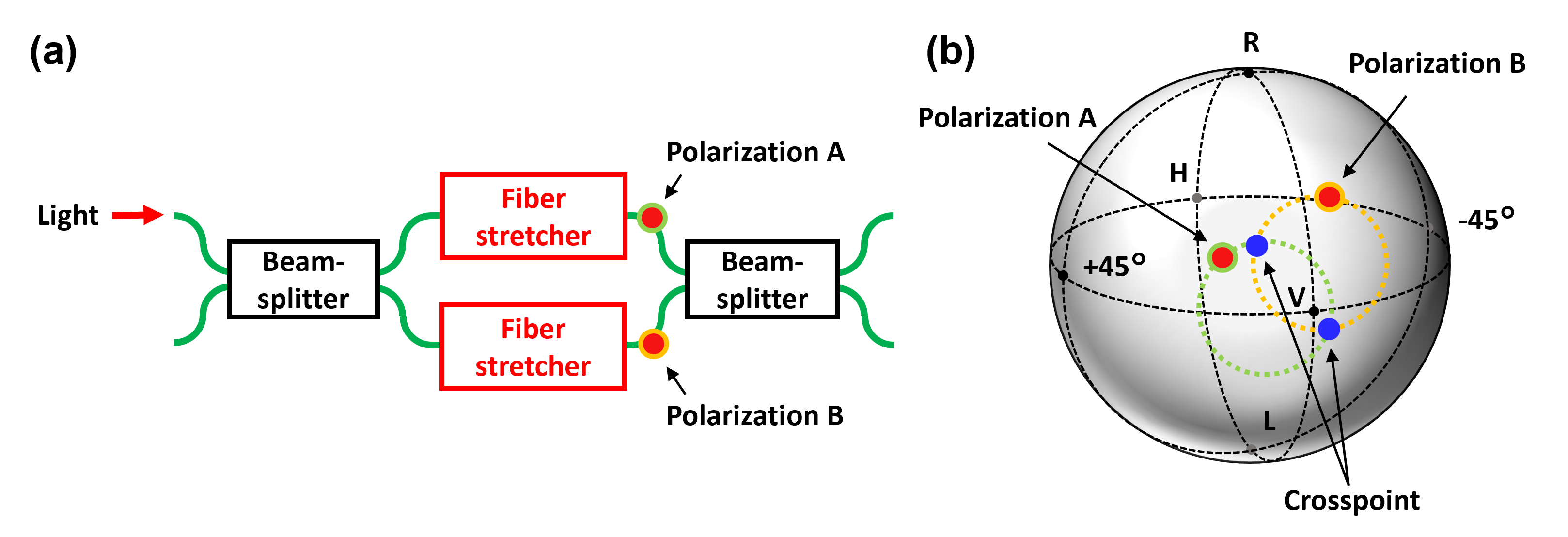}
\caption{Principle of the CCC method. (a) Interferometer with fiber stretchers inserted. (b) Two circular trajectories drawn by polarization A and B on the Poincar\'{e} sphere when the voltages applied to the piezo actuators of the fiber stretchers are varied. In the CCC method, polarization states are controlled to a crosspoint of the two circles to optimize visibility. Poincar\'{e} sphere is explained in Section \ref{sec2}. V, H, R and L correspond to vertical, horizontal, right-handed circular and left-handed circular polarizations, respectively.}
\label{fig1}
\end{figure}

In this paper, we propose a method to optimize visibility in a fiber interferometer by controlling the polarization to a crosspoint of two circular trajectories drawn on a Poincar\'{e} sphere (Fig.~\ref{fig1}). We also experimentally demonstrate visibility maximization with this method, where the visibility is maintained basically above 99.9\% for three hours. The circular trajectories appear when the two light beams are off the polarization-maintaining axis of polarization-maintaining fibers and when the fibers are stretched or heated. We call this the circle-circle crosspoint (CCC) method in the following. The CCC method is especially suitable for the quantum light because it can be implemented with low losses. The CCC method can be implemented with fiber stretchers or heaters, which do not cause severe microbends resulting in optical losses. Another important point is that, by employing the CCC method, fiber systems can be constructed solely with the polarization-maintaining fibers. In contrast, the conventional methods employ non-polarization-maintaining fibers for the polarization controlling parts, while other parts are constructed with polarization-maintaining fibers. Because there is no connection between the polarization-maintaining fibers and the non-polarization-maintaining fibers, it is expected that the optical losses tend to be lower. For the experimental demonstration, we made fiber stretchers and measured the optical losses, which were as low as 0.02~\text{dB} (0.5\%). As a result, the visibility in three hours was basically maintained above 99.9\% with the polarization control, while it sometimes dropped to 98\% without the polarization control. Although this experiment is demonstrated for a single Mach-Zehnder interferometer, the CCC method can be applied to more complex interferometers having multiple interference points by maximizing visibilities from the upstream to the downstream of the light paths.

Section \ref{sec2} describes the Poincar\'{e} sphere representation of the polarization states. Section \ref{sec3} discusses the change in polarization states at the output of polarization-maintaining fibers. Section \ref{sec4} illustrates the effect of a fiber stretcher on polarization. Section \ref{sec5} shows low-loss polarization control using fiber stretchers. In Section \ref{sec6}, the fabrication method of the fiber stretchers for polarization controllers are described. In Section \ref{sec7}, the experimental method for polarization control using the CCC method is described. Section \ref{sec8} demonstrates the experimental result. Section \ref{sec9} shows the comparison of the CCC method and the conventional methods. Section \ref{sec10} presents the conclusion of this study.

\section{Poincar\'{e} sphere representation} \label{sec2}
\begin{figure}[tb]
\centering
\includegraphics[width=\linewidth]{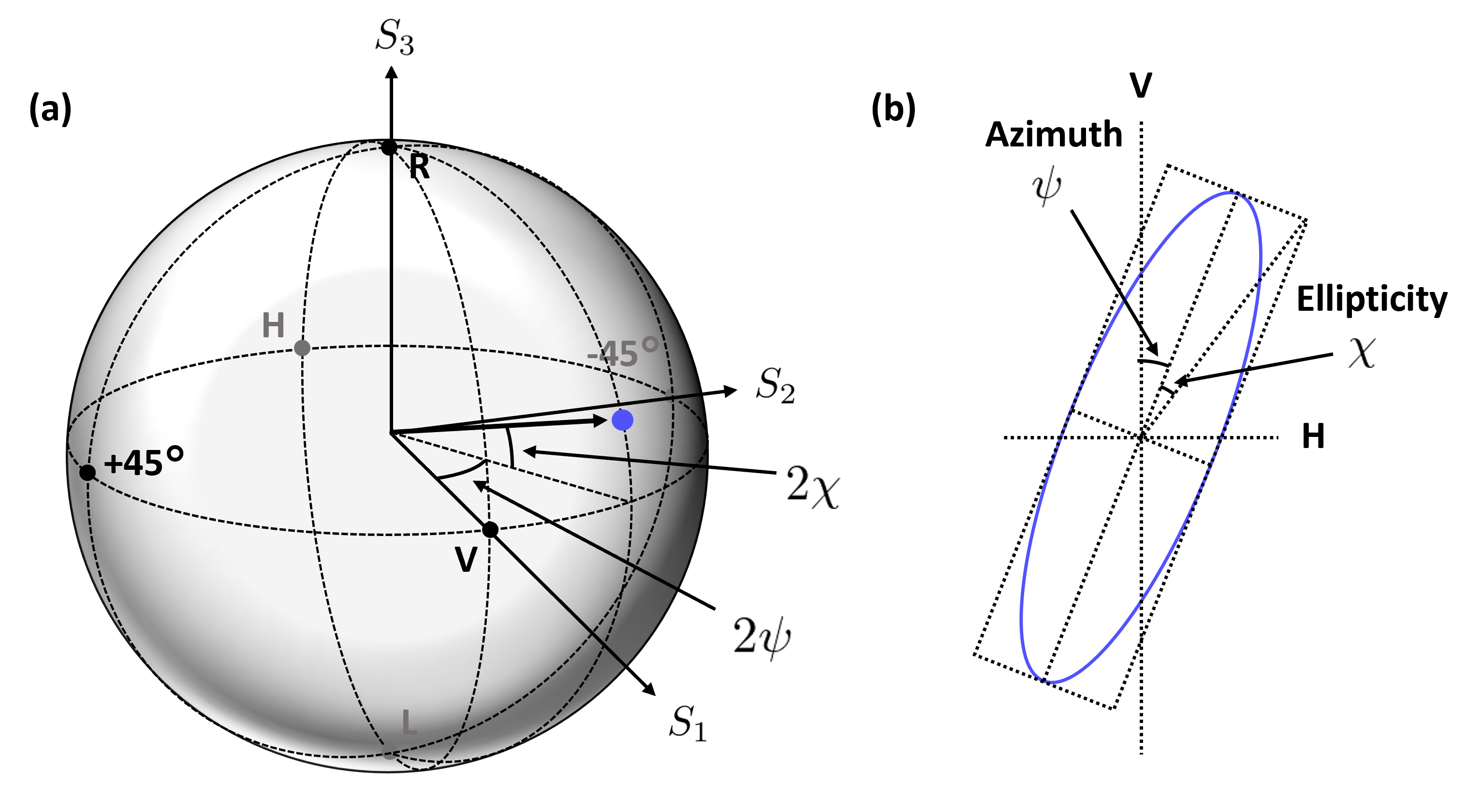}
\caption{Polarization states. (a) Poincar\'{e} sphere. (b) The azimuth $\psi$ and the ellipticity $\chi$ in elliptical polarization states.}
\label{fig2}
\end{figure}

We introduce the Poincar\'{e} sphere, on which we can describe polarization states. Let $x$ and $y$ components of the electric field of monochromatic light propagating in the $z$ direction be
\begin{subequations}\label{eqn1}
\begin{align}
E_x &= a_x \cos(kz - \omega_c t - \Gamma_x),\label{eqn1a}\\
E_y &= a_y \cos(kz - \omega_c t - \Gamma_y),\label{eqn1b}
\end{align}
\end{subequations}
where $a_x$ and $a_y$ are $x$ and $y$ components of the electric field amplitudes, $\omega_c$ is the angular frequency of light, $k$ is the wave number, and $\Gamma_x$ and $\Gamma_y$ are the phases of $x$ and $y$ components. Stokes parameters are defined as follows:
\begin{subequations}\label{eqn2}
\begin{align}
S_0 &= a_x^2+a_y^2, \label{eqn2a}\\
S_1 &= a_x^2-a_y^2, \label{eqn2b}\\
S_2 &= 2a_x a_y \cos\Gamma, \label{eqn2c}\\
S_3 &= 2a_x a_y \sin\Gamma, \label{eqn2d}
\end{align}
\end{subequations}
where $\Gamma=\Gamma_y-\Gamma_x$. Since $S_0$ corresponds to the optical power and $S_1^2+S_2^2+S_3^2 = S_0^2$, a set of vectors $(S_1, S_2, S_3)$ forms a sphere called the Poincar\'{e} sphere when the optical power is constant (Fig.~\ref{fig2}(a))\cite{Johnson1981}. Also, Stokes parameters $S_1, S_2, S_3$ are expressed as follows:
\begin{subequations} \label{eqn3}
\begin{align}
S_1 &= S_0 \cos 2\chi \cos 2\psi, \label{eqn3a}\\
S_2 &= S_0 \cos 2\chi \sin 2\psi, \label{eqn3b}\\
S_3 &= S_0 \sin 2\chi, \label{eqn3c}
\end{align}
\end{subequations}
where $\psi$ is the azimuth and $\chi$ is the ellipticity  \cite{Born2019}. For elliptical polarization states, the azimuth $\psi$ and the ellipticity $\chi$ are expressed in Fig.~\ref{fig2}(b).

\section{Polarization states through polarization-maintaining fibers} \label{sec3}
\begin{figure}[tb]
\centering
\includegraphics[width=\linewidth]{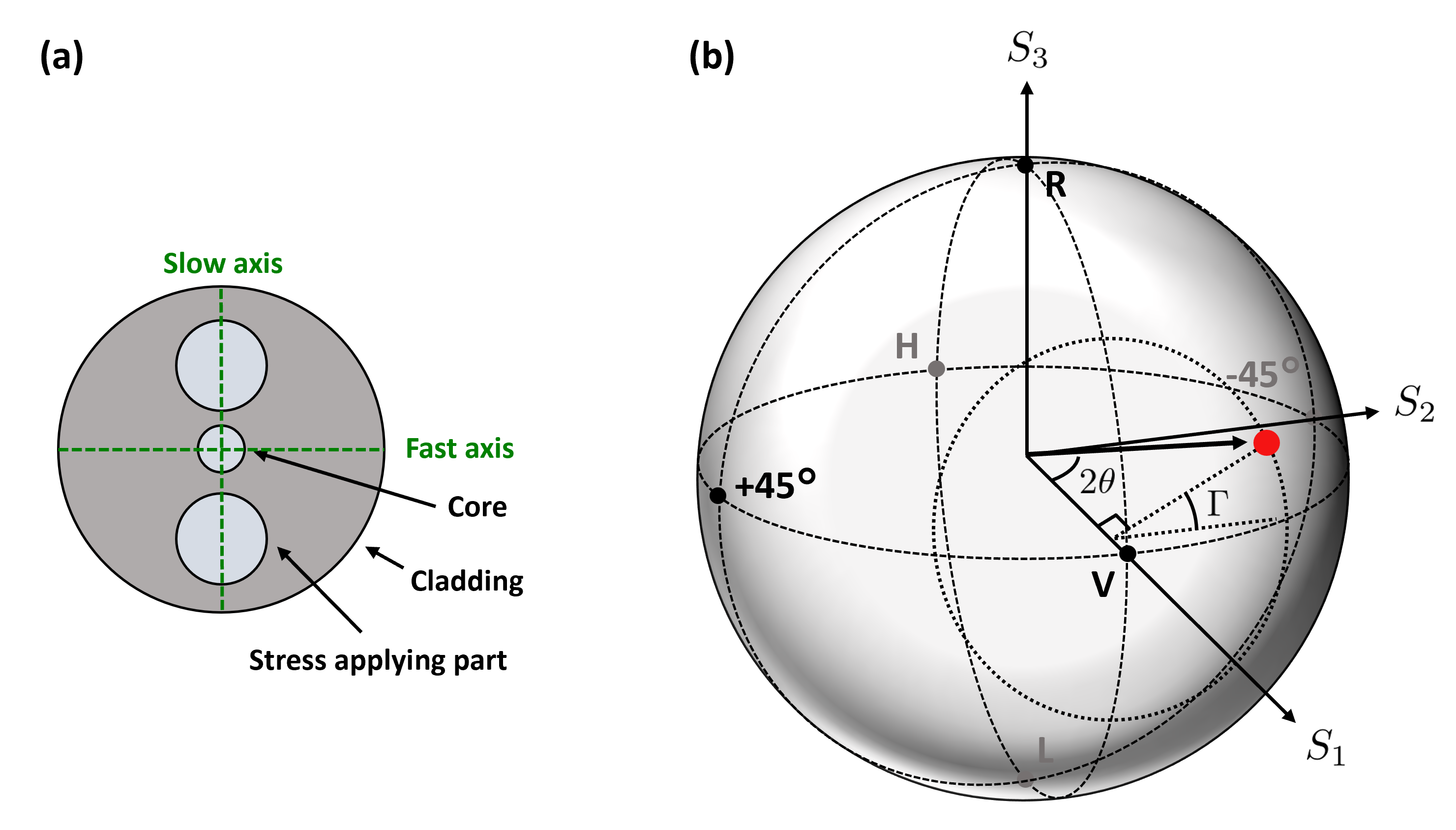}
\caption{Polarization-maintaining fibers and polarization changes. (a) Cross section of polarization-maintaining fiber. (b) $\Gamma$ and $\theta$ in the Poincar\'{e} sphere.}
\label{fig3}
\end{figure}

The polarization-maintaining fiber has a core and cladding with a stress-applying part to generate a birefringence as shown in Fig.~\ref{fig3}(a), and realizes an anisotropic refractive index distribution. The direction in which the stress-applying part exists is called the slow axis, and the direction orthogonal to the slow axis is called the fast axis. In this paper, we call the linear polarization state parallel to the slow axis as the polarization state of the point V on the Poincar\'{e} sphere. When the linearly polarized input light is parallel to either the slow axis or the fast axis, the polarization is maintained through the fiber. Now we consider the output polarization state from the polarization maintaining fiber when the input linear polarization state is $\theta$ rotated from the slow axis. Because the slow-axis component and the fast-axis component undergo different phase changes, the electric field of the output light from a fiber, corresponding to Eq.~(\ref{eqn1}), becomes,
\begin{subequations} \label{eqn4}
\begin{align}
E_x &= a \cos\theta \cos(kz - \omega_c t - \Gamma_x), \label{eqn4a}\\
E_y &= a \sin\theta \cos(kz - \omega_c t - \Gamma_y), \label{eqn4b}
\end{align}
\end{subequations}
where $a^2=a_x^2+a_y^2$. In this case, the Stokes parameters Eq.~(\ref{eqn2}) are
\begin{subequations} \label{eqn5}
\begin{align}
S_0 &= a^2, \label{eqn5a}\\
S_1 &= a^2 \cos(2\theta), \label{eqn5b}\\
S_2 &= a^2 \sin(2\theta) \cos\Gamma, \label{eqn5c}\\
S_3 &= a^2 \sin(2\theta) \sin\Gamma. \label{eqn5d}
\end{align}
\end{subequations}
In the Poincar\'{e} sphere, $\theta$ and $\Gamma$ are depicted in Fig.~\ref{fig3}(b). 
The phase difference $\Gamma$ changes depending on the length of the fiber and the refractive index difference between the slow axis and the fast axis, while $\theta$ does not change. The fiber length where $\Gamma$ changes by $2\pi$ is called the beat length. The beat length is about 4 mm with ordinary fibers for the communication wavelength band\cite{FujikuraPM}. Changes in the temperature or stress applied to the fiber slightly alter the fiber length and the refractive index difference, resulting in changes in the phase difference $\Gamma$ at the fiber exit. With this changes of $\Gamma$, ($S_2, S_3$) draws a circular trajectory from Eq.~(\ref{eqn5c}) and Eq.~(\ref{eqn5d}). The radius of the circle depends on $\theta$. At $\theta=0$, the trajectory of the polarization state converges to the point V in the Poincar\'{e} sphere and the polarization is no longer affected by the disturbance.

\begin{figure}[tb]
\centering
\includegraphics[width=\linewidth]{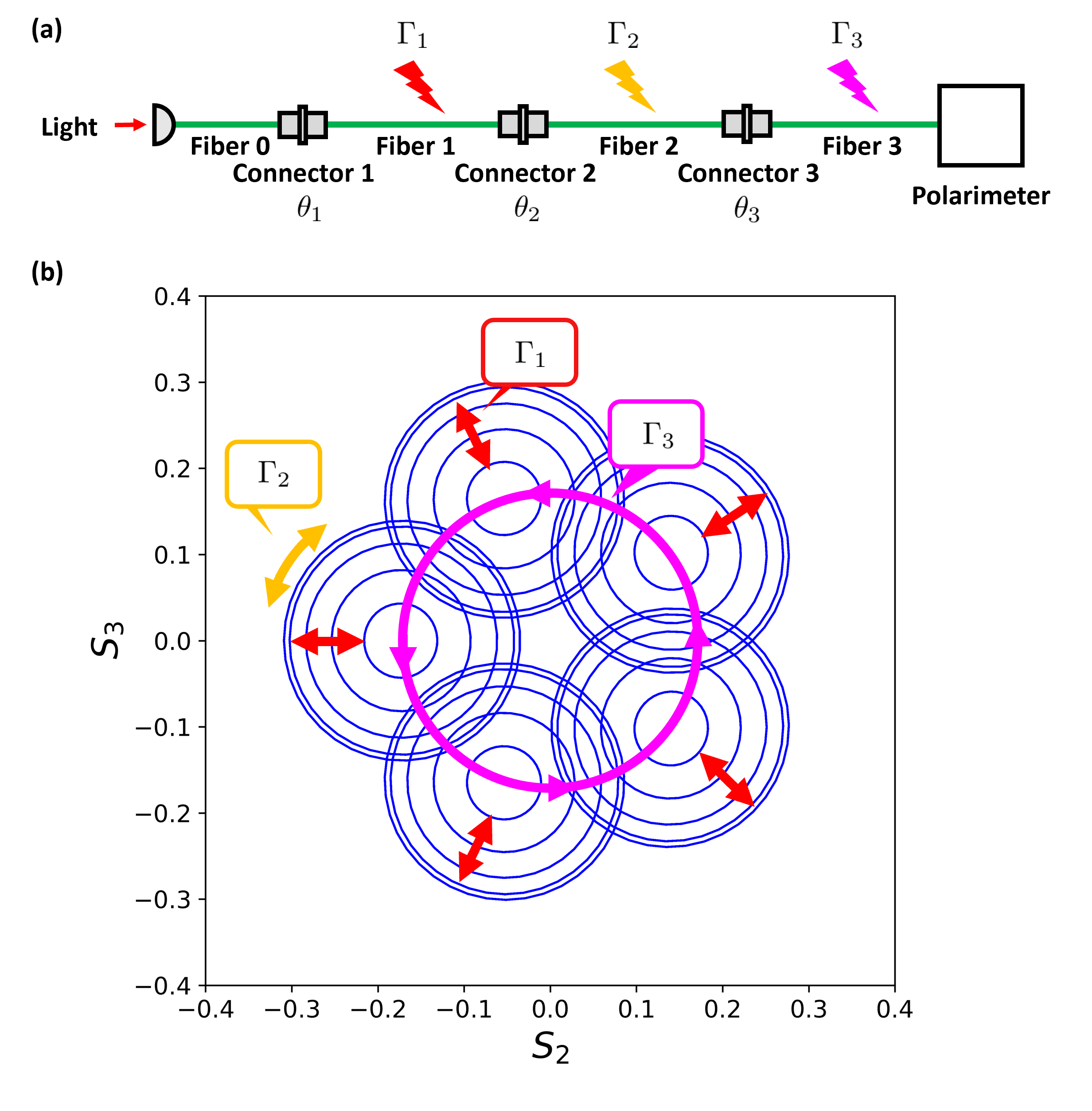}
\caption{Polarization changes through polarization-maintaining fibers with multiple connectors. (a) Setup with misalignments $\theta_1$, $\theta_2$, $\theta_3$ at connectors. (b) Circular trajectories when a disturbance of $2\pi$ is added to $\Gamma_2$. Here $S_0=1$. A slight modification in $\Gamma_1$ changes the radius of the circle, and a slight change in $\Gamma_3$ shifts the center of the circle.}
\label{fig4}
\end{figure}

Now we consider the polarization state of light propagating through multiple polarization-maintaining fibers connected by connectors with misalignments as shown in Fig.~\ref{fig4}(a). Suppose that the light propagating through fiber 0 is in a linearly polarized state and the polarization plane is coincident with the polarization-maintaining axis.
Fiber 0 is sequentially connected to Fiber 1, Fiber 2, and Fiber 3  by Connector 1, Connector 2, and Connector 3, respectively. The angles between slow axes of two connected fibers at Connector 1, Connector 2, and Connector 3 are expressed by $\theta_1, \theta_2, \theta_3$, respectively. Although connectors are supposed here, in actual situations, there are various components that cause polarization-axis mismatches, such as incomplete fiber fusion splices. Let $\Gamma_1$, $\Gamma_2$, and $\Gamma_3$ be the phase differences between the slow axis and the fast axis which light receives through Fiber 1, Fiber 2, and Fiber 3, respectively. When disturbances are added to Fiber 1, Fiber 2, and Fiber 3, $\Gamma_1$, $\Gamma_2$, and $\Gamma_3$ shift, respectively. The output light from Fiber 3 is measured with a polarimeter. The Poincar\'{e} sphere for the output light is defined so that the slow axis of Fiber 3 corresponds to the point V. When any of $\Gamma_1$, $\Gamma_2$, and $\Gamma_3$ changes, the polarization state draws a circular trajectory on the Poincar\'{e} sphere. In particular, in Fig.~\ref{fig4}(b), the circular trajectory by changing $\Gamma_2$ more than $2\pi$ is shown by a blue circle for various $\Gamma_1$ and $\Gamma_3$. In this figure, $\theta_1$, $\theta_2$, and $\theta_3$ are set to 2, 2, and 5 degrees for the sake of visual understandability. These are typical amounts of misalignments when connected by connectors. When $\Gamma_1$ is slightly altered, the radius of the circle changes. When $\Gamma_3$ is slightly altered, the center of the circle changes. Thus, the radius and the center point of the circle can be changed by adding disturbances before and after the fiber.

\section{Phase and Polarization changes by fiber stretchers} \label{sec4}
Fiber stretchers pull fibers to change the optical path lengths which are typically used for controlling the phases of light. A fiber stretcher has a structure, for example, where a fiber is wound around a cylindrical object with a variable radius. The radius of the cylindrical object is typically changed by a piezo actuator, which is controlled electrically. Other phase control devices for fiber systems include Electro-Optical Modulator (EOM)\cite{Sinatkas2021}, which uses the electro-optic effect. EOMs have the advantage of a wide bandwidth of tens of gigahertz. However, fiber EOMs have the disadvantage of severe optical losses of typically 3~\text{dB} (50\%)\cite{EOSPACE}. In contrast, fiber stretchers have a low bandwidth of about 100~\text{kHz}, which is determined by the bandwidth of the piezo actuator. However, it has the advantage that the optical loss is typically less than 0.2~\text{dB} (4.5\%)\cite{LUNA_fst}. Fiber stretchers are chosen when phase controls are needed for quantum light, which is vulnerable to optical losses. In optical quantum computers, quantum light is combined with classical light for measurement. If appropriate ancillary states can be prepared\cite{Gottesman2001, Konno2021}, universal quantum computation will be realized by fast measurement switching on the cluster state\cite{Menicucci2006, Asavanant2021, Larsen2021a}. Fast phase controls, therefore, are only needed for classical light, which is done by EOMs, and slow phase controls are sufficient for quantum light, which is done by fiber stretchers.

When a polarization-maintaining fiber is stretched by a fiber stretcher, the phase difference $\Gamma$ between the slow and fast axes changes. If the polarization state of the input light, therefore, does not coincide with the polarization-maintaining axes, the polarization state changes, and draws a circular trajectory on the Poincar\'{e} sphere as discussed in Section \ref{sec3}. Note that similar polarization changes can be also induced by temperature changes using such as heaters. We will describe in Section \ref{sec5} how to use these polarization changes for maximizing visibility. 

\section{Maximizing visibility by the circle-circle crosspoint method} \label{sec5}

\begin{figure}[tb]
\centering
\includegraphics[width=\linewidth]{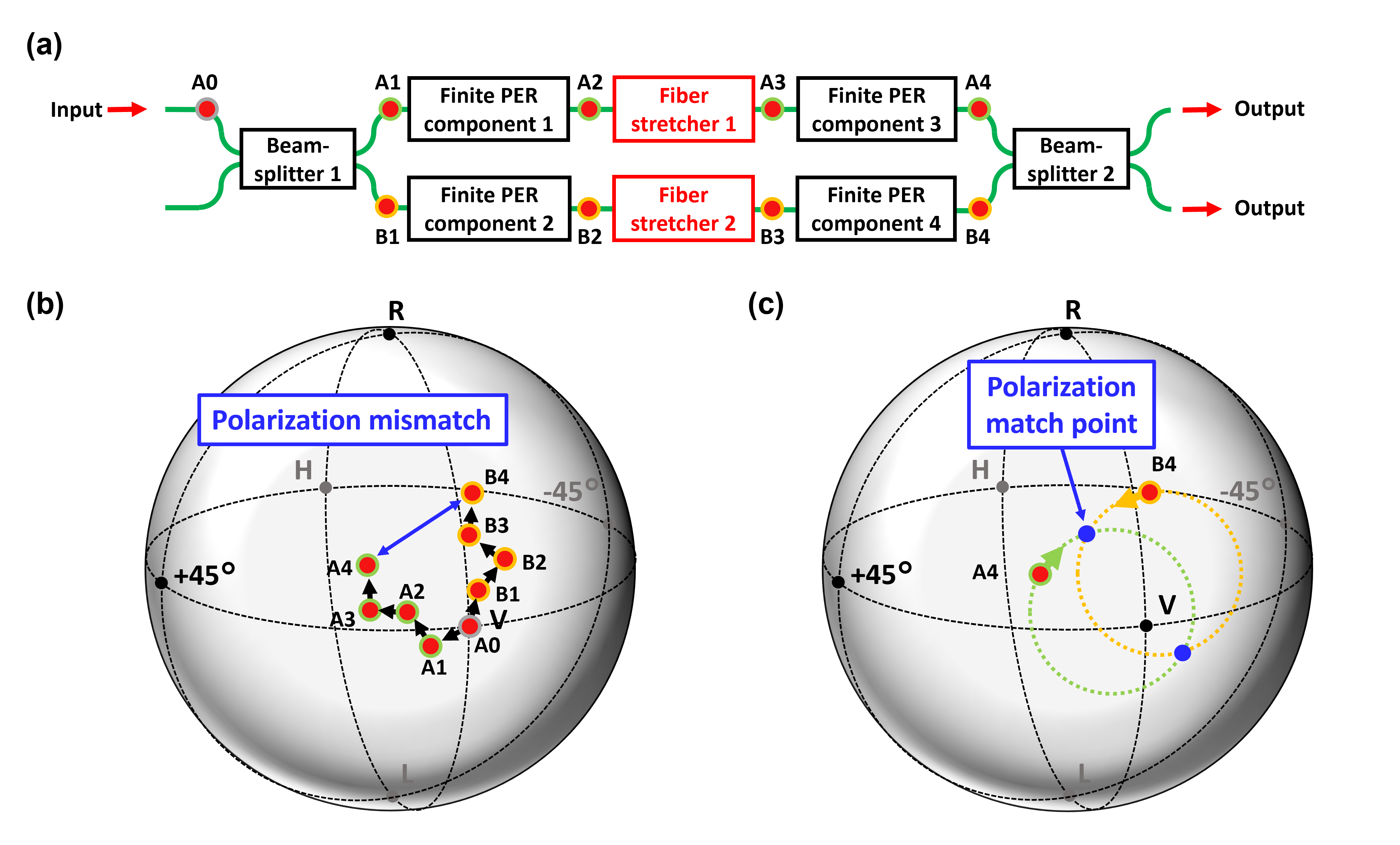}
\caption{The CCC method in an interferometer with multiple finite PER components. (a) An interferometer setup for considering the changes in polarization states. (b) Polarization changes on the Poincar\'{e} sphere when light passes through the finite PER components in the interferometer. (c) Two circular trajectories drawn by A4 and B4 when voltages applied to Fiber stretcher 1 and Fiber stretcher 2 are changed. In the CCC method, A4 and B4 are controlled to a crosspoint of the two circles.}
\label{fig5}
\end{figure}

So far, we have considered a single optical path system with multiple fiber elements connected. In this section, we will consider the case of an optical interferometer as shown in Fig.~\ref{fig5}(a). The Mach-Zehnder interferometer consists of two beamsplitters. In fiber systems, these beamsplitters have finite PERs (typically 25~\text{dB} or more). In addition, these paths of the interferometer often have fiber stretchers for phase locking and tapping fiber beamsplitters for monitoring the state of the light, which also have finite PERs. Furthermore, when connecting these fiber elements, if the fibers are fused together, the PER of fusion splices is typically about 35~\text{dB}, and if they are connected with connectors, there are typically axis misalignments of 2 degrees (PER = 29~\text{dB}) to 5 degrees (PER = 21~\text{dB}).

The interferometer has two optical paths, and the polarization state of each optical path changes independently as shown in Fig.~\ref{fig5}(b). We suppose that the polarization state A0 of the input to the interferometer is at point V on the Poincar\'{e} sphere. As the light passes through Beamsplitter 1, the light splits into two paths, but due to the finite PER of the fiber beamsplitter, the polarization state shifts slightly to A1 and B1 on the Poincar\'{e} sphere, respectively. Then, due to the finite PER of the elements in each path, the polarization state of one changes from A1 to A2, A3, and A4, while the other changes from B1 to B2, B3, and B4. When light enters Beamsplitter 2, the visibility decreases as the distance between polarization states A4 and B4 on the Poincar\'{e} sphere increases.

In the CCC method, both paths of the interferometer have fiber stretchers (or heaters) for polarization control. By driving the fiber stretchers, the polarizations of A4 and B4 draw two different circular trajectories as shown in Fig.~\ref{fig5}(c). In most cases, these two circular trajectories have crosspoints. By adjusting the fiber stretchers to the crosspoint of these circles, polarization states A4 and B4 are matched. Visibility, therefore, is maximized by this polarization control.

If the circular trajectories do not have any crosspoint, visibility does not reach 100\%. However, as explained in Section \ref{sec4}, the radius and the center of the circle trajectory can be moved by placing a device that causes a phase difference between the slow and fast axes before and after the fiber stretcher is inserted. The trajectories of the two circles, therefore, can be changed to have crosspoints.

If the polarization control is performed by heaters or Peltier devices instead of fiber stretchers, larger dynamic ranges are realized. This larger dynamic range may be useful in some situations.

The bandwidth of the fiber stretcher is typically up to about 100~\text{kHz}, and heaters or Peltier devices are much slower. But since the polarization control for visibility maximization only needs to follow the slow drift of polarization, high speed is not necessary and these slow polarization controllers are sufficient to achieve visibility maximization.

\begin{figure}[tb]
\centering
\includegraphics[width=0.8\linewidth]{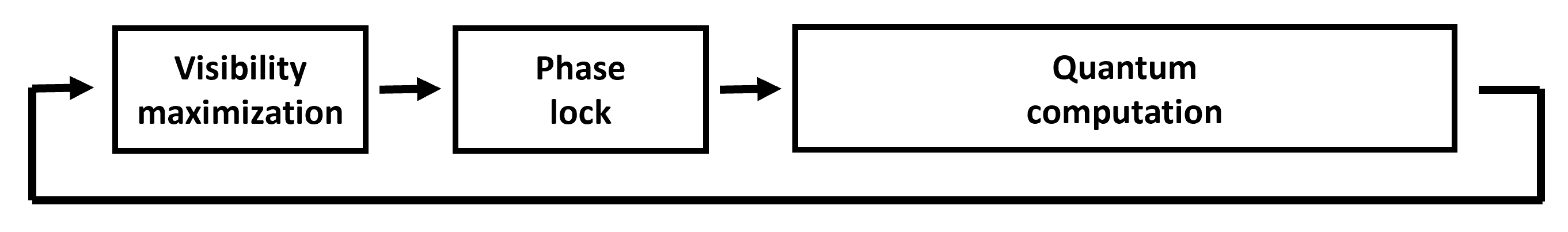}
\caption{Processes for operating a quantum computer.}
\label{fig6}
\end{figure}

In addition to visibility maximizations, phase lock is necessary for optical quantum information processing. However, polarization control in the CCC method causes a significant phase drift. Polarization control and phase control, therefore, should be performed in two steps. Firstly, polarization control is performed with phase control off, and the polarization controllers are held at the optimal polarization states. Secondly, the phase control is turned on. Phase control can be performed by connecting a small fiber stretcher in series with the polarization controller. Optical quantum information processing is performed after the preparation by this two-step control (Fig.~\ref{fig6}). In principle, it is also possible to perform polarization control and phase control in a single fiber stretcher. This is because the voltage scales applied to the fiber stretcher for polarization control and phase control are significantly different. It is, therefore, possible to first perform polarization control on a larger voltage scale and then perform phase control on a smaller voltage scale.

In the CCC method, optimized polarization states are shifted from the slow axis of the polarization-maintaining fiber. Hence, the polarization dependence of the beamsplitter becomes the nonideality of the interferometer, which degrades the quality of the quantum computation. In order to assess these problems, we conducted some measurements. When various polarizations of light are input to a fiber beamsplitter 954P (Evanescent optics) with a coupling ratio of 49:51, the variation of the coupling ratio was $\pm$0.3\%, which is usually negligible.

Although the CCC method is explained for a simple Mach-Zehnder interferometer, the CCC method can be applied to systems that have multiple interference points, by maximizing visibilities from the upstream to the downstream of the system.

\section{Making  of fiber stretchers} \label{sec6}
The fiber stretchers for the CCC method should be able to change the phase difference between the slow and fast axes larger than $2\pi$. Due to the small difference of refractive indices between slow and fast axes of the polarization-maintaining fiber, shifts of the phase difference by $2\pi$ empirically correspond to a few hundred wavelengths of phase shifts. Hence, larger fiber stretchers are required in comparison with those for phase control which changes the phase by a few wavelengths.

\begin{figure}[tb]
\centering
\includegraphics[width=\linewidth]{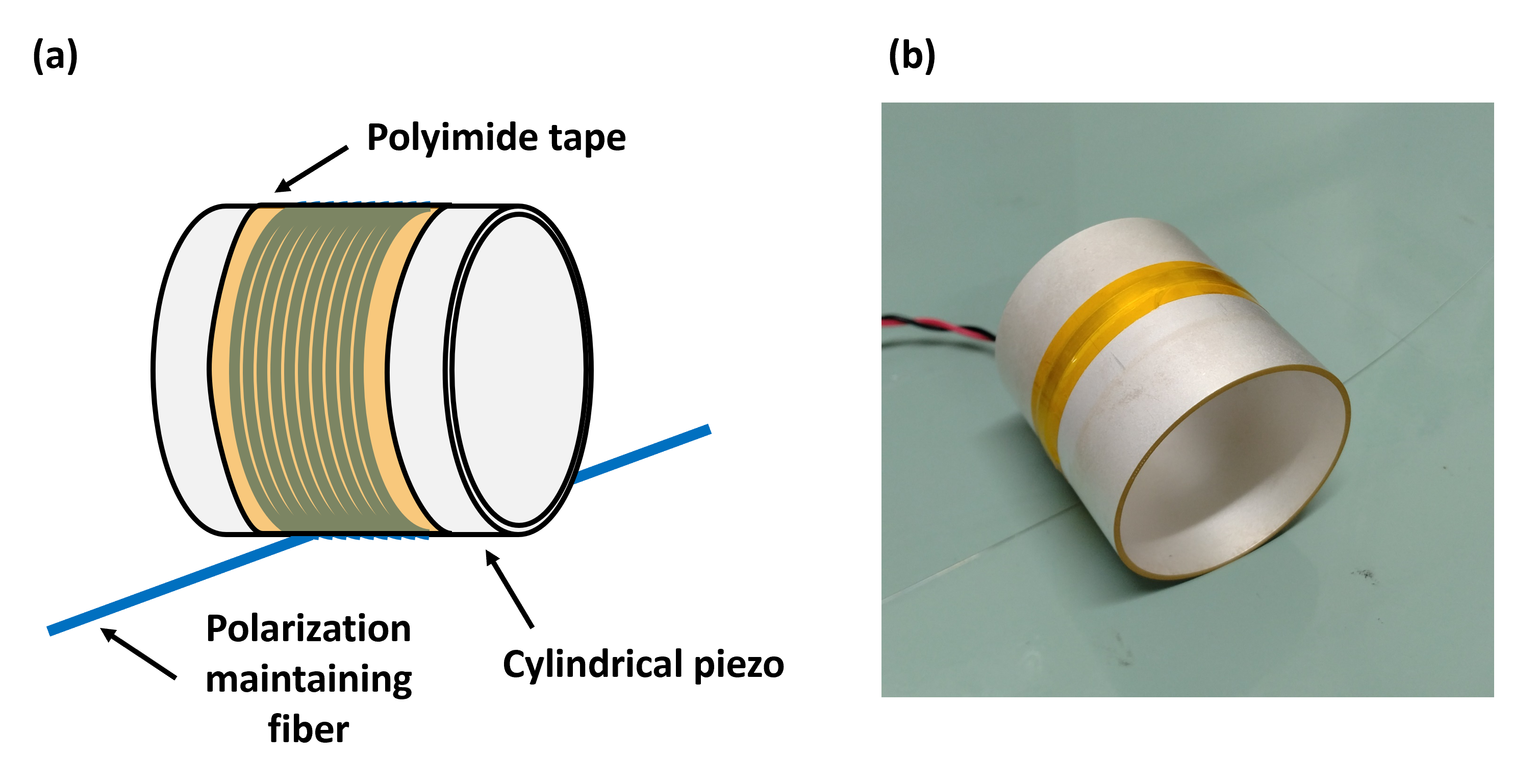}
\caption{Fiber stretcher for polarization control. (a) Structure of the fiber stretcher. (b) Photo of the fiber stretcher.}
\label{fig7}
\end{figure}

Although fiber stretchers are commercially available, in this study we made our own fiber stretchers because we wanted to customize the number of fiber windings and the radius at which the fibers are wound. The fiber stretcher we fabricated utilizes a 40 mm diameter cylindrical piezo actuator PT140.70 (PI)\cite{PIjapan}, around which a polarization-maintaining fiber for communication wavelengths SM15-PS-U25D (Fujikura) is wrapped about 10 times, as shown in Fig.~\ref{fig7}. The wound fiber is fixed with Polyimide tape. A voltage of up to 1000~\text{V} can be applied to the piezo actuator and a maximum diameter contraction is 12 um. Hence, if all the piezoelectric contraction contributed to fiber stretching, the fiber would be stretched by 0.37 mm. The fiber stretcher we fabricated draws a single circle at about 600~\text{V}, which corresponds to about 7 um contraction. This contraction causes the change of the fiber length by about 200 um, which is much smaller than the beat length about 4 mm of the polarization-maintaining fiber\cite{Kaminow1981, FujikuraPM}. Hence, the change of the phase difference is mainly due to stress-induced birefringence, rather than the change in the fiber length.

The measured insertion loss of the fiber stretcher was 0.02~\text{dB} (0.5\%), which is acceptable for quantum light. Furthermore, this optical loss is almost unchanged when the voltage applied to the fiber stretcher is changed from 0~\text{V} to 1000~\text{V}.

\begin{figure}[tb]
\centering
\includegraphics[width=0.7\linewidth]{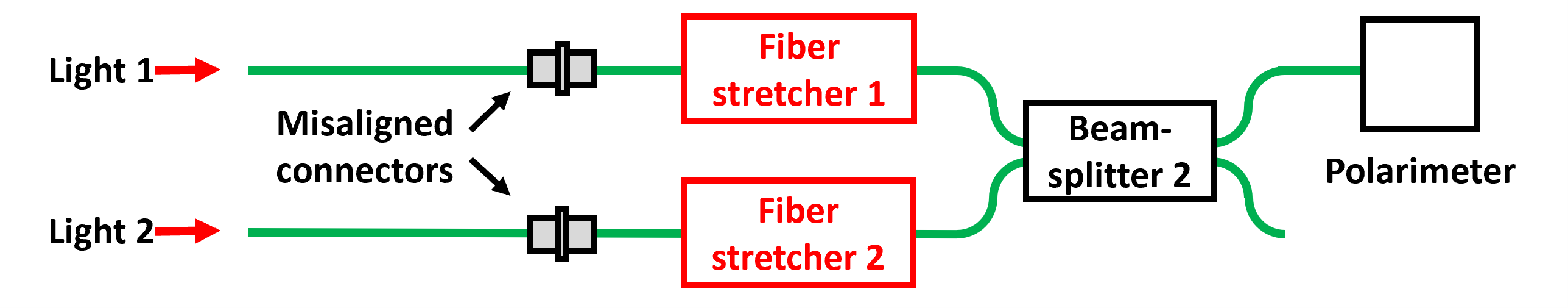}
\caption{Setup for the polarimeter measurement of the circular trajectories of polarization drawn on the Poincar\'{e} sphere. }
\label{fig8}
\end{figure}

\begin{figure}[tb]
\centering
\includegraphics[width=0.7\linewidth]{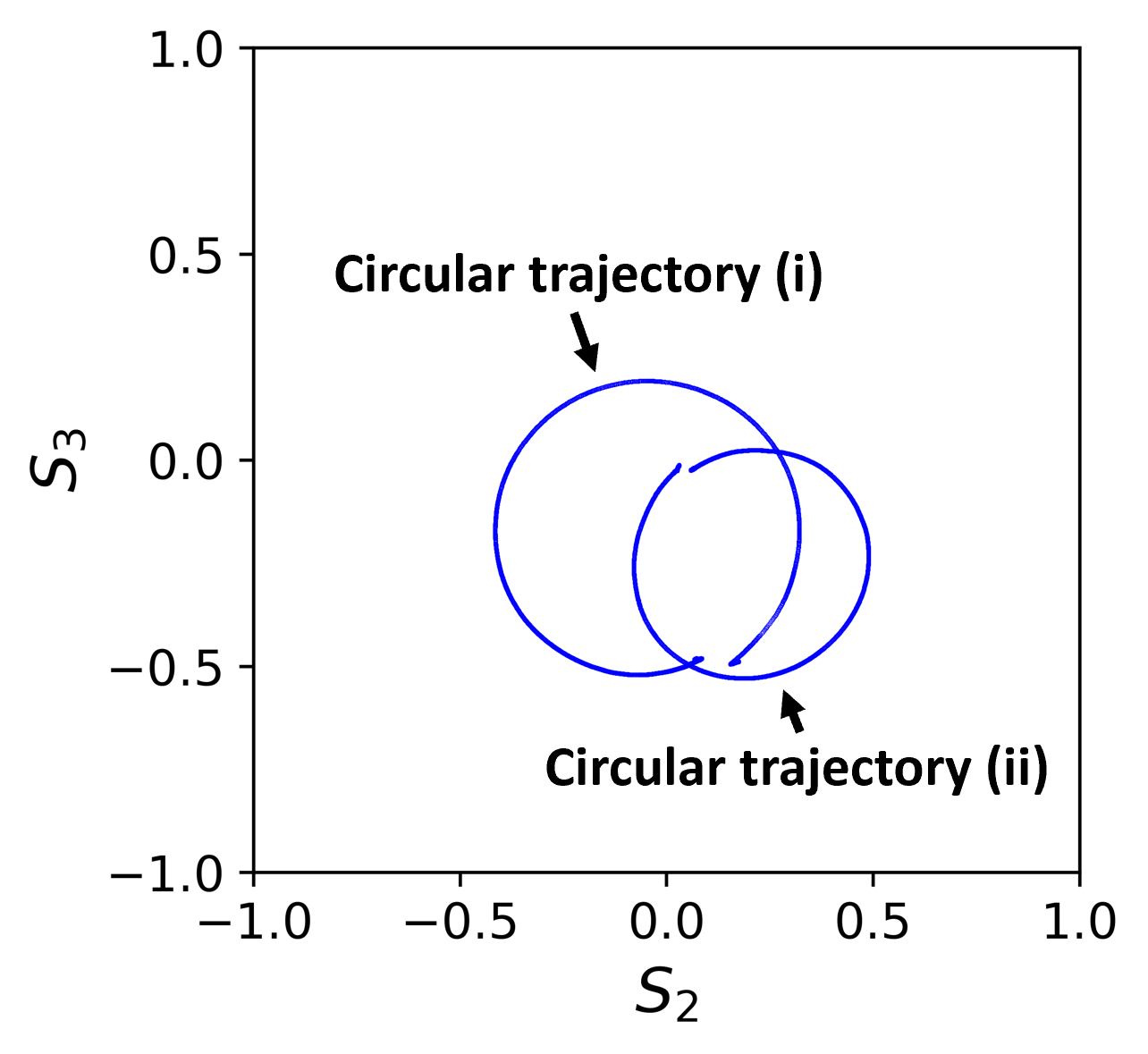}
\caption{Two circular trajectories obtained by the setup in Fig.~\ref{fig8}, which have cross points. Here $S_0=1$.}
\label{fig9}
\end{figure}

The actual circular trajectories are measured by the setup in Fig.~\ref{fig8} and obtained polarization trajectories are shown in Fig.~\ref{fig9}. Firstly, only Light 1 was fed and the voltage applied to Fiber stretcher 1 was changed, resulting in Circular trajectories (i) in Fig.~\ref{fig9}. Secondly, only Light 2 was fed and the voltage applied to Fiber stretcher 2 was varied, resulting in Circular trajectories (ii) in Fig.~\ref{fig9}. In the CCC method, polarizations are controlled to the crosspoints of the two trajectories. Note that these circular trajectories are not obtained from the experimental setup in Sec.~\ref{sec7}, where all fiber components are connected by fusion splices and thus we cannot see the polarization trajectory of each optical path like Fig.~\ref{fig9}.

\section{Experimental Method} \label{sec7}
\begin{figure}[tb]
\centering
\includegraphics[width=\linewidth]{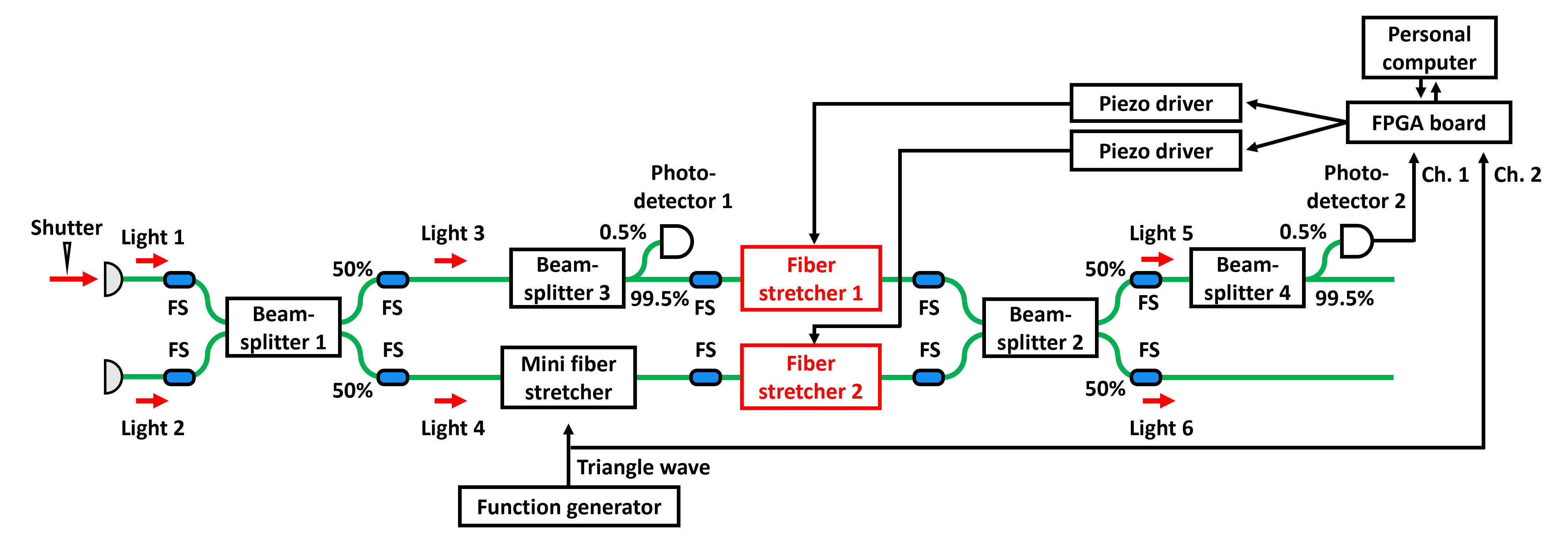}
\caption{Experimental setup for visibility maximization with the CCC method using Fiber stretcher 1 and Fiber stretcher 2. FS: fusion splice.}
\label{fig10}
\end{figure}

We experimentally demonstrate visibility maximization by the CCC method with the setup shown in Fig.~\ref{fig10}. The polarization state of Light 1 is adjusted as much as possible so that the light is linearly polarized to match the slow axis of the polarization-maintaining fiber. Light 1 is split into Light 3 and Light 4 by passing through Beamsplitter 1, which has a coupling ratio of 50:50. Light can also be input from the Light 2 path, although Light 2 is blocked in this experiment. Light 3 and Light 4 interfere at Beamsplitter 2 with a coupling ratio of 50:50 to generate Light 5 and Light 6. Beamsplitter 3 with the coupling ratio of 0.5:99.5 is inserted in the path of Light 3 so that the interference phase between Light 1 and Light 2 can be monitored by Photodetector 1 and locked by feedback control. A mini fiber stretcher is inserted in the Light 4 path to scan the phase, by which an interference signal between Light 3 and Light 4 is generated to calculate the visibility. In both the Light 3 and Light 4 paths, Fiber stretcher 1 and Fiber stretcher 2, which are explained in Section \ref{sec6}, are inserted to control the polarization. To monitor the interference signal, Light 5 is input to Beamsplitter 4 with the coupling ratio of 0.5:99.5, and the output of 0.5\% is detected by Photodetector 2. Light 1 has an optical power of about 1.2 mW. Beamsplitter 1, Beamsplitter 2, Beamsplitter 3, and Beamsplitter 4 are 954P (Evanescent optics). The individual difference in the coupling ratio for a 50:50 beamsplitter is $\pm$2\%. The voltages are applied to Fiber stretcher 1 and Fiber stretcher 2 with a piezo actuator driver, SVR 1000/3 (piezo actuatormechanik GmbH), which can apply up to 1000~\text{V}. The mini fiber stretcher is FS20 (IDIL). With this mini fiber stretcher, the phase can be scanned by about 5 wavelengths in the voltage range of 0~\text{V} to 70~\text{V}. Triangle wave signal with a frequency of 100 Hz is applied to this mini fiber stretcher. The triangle wave signal is generated by an arbitrary wave generator (33500B Keysight) whose ±0.8~\text{V} signal is amplified to the range between 20~\text{V} and 50~\text{V} by a homemade amplifier. Photodetector 2 is homemade and the photodiode is G8195-11 (Hamamatsu Photonics). The transimpedance gain is 100 k$\Omega$, the reverse bias is 15~\text{V}, and the flat bandwidth is about 30~\text{MHz}. A motor-driven optical shutter is placed in the Light 1 path to periodically block the light and retake the voltage level of Photodetector 2 without light input, which may drift during the experiment.

Interference voltage signals which are obtained by Photodetector 2 are sent to analog-to-digital converters on an FPGA board, STEMlab 125-14 (RedPitaya). The FPGA board has an analog input voltage range of ±1V, a voltage resolution of 14 bits, and a sampling frequency of 125~\text{MHz}. The data are then downsampled to 1/256 in the FPGA board and finally acquired at a sampling frequency of 488~\text{kHz}. In the downsampling, 256 points are averaged in the FPGA board with 14-bit accuracy.

\begin{figure}[tb]
\centering
\includegraphics[width=\linewidth]{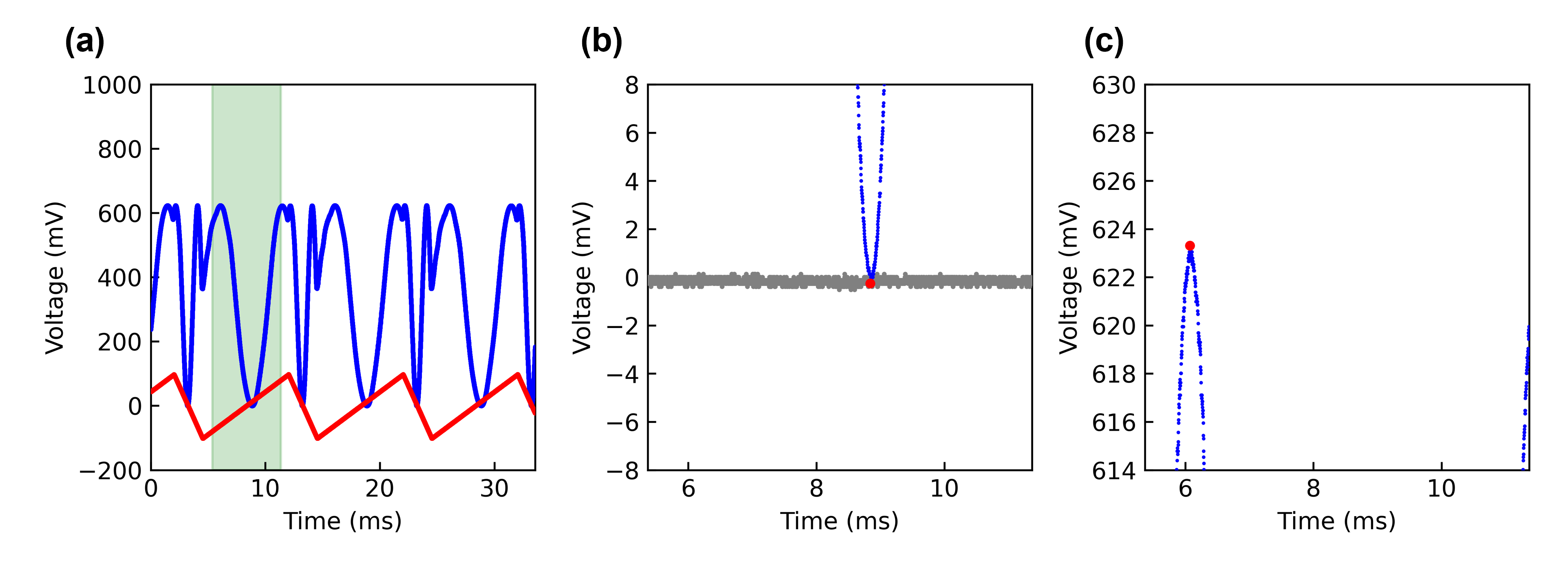}
\caption{Example of an interference signal to obtain visibility. (a) Interference signal acquired in Ch.~1 of the FPGA board (blue points) and triangle wave signal applied to the mini fiber stretcher acquired in Ch.~2 (red points). Min and max of the interference signal obtained in the green area, where the triangle wave signal monotonically increases. (b) Magnified interference signal (blue points) around the min (red point), together with the detector signal without light (gray points) as a reference. (c) Magnified interference signal (blue points) around the max (red point). }
\label{fig11}
\end{figure}

The digital data acquired on the FPGA board is transferred to a personal computer. Data of 16384 points are acquired at one time on the FPGA board, which we call one frame of data. The time width of the one frame is 33~\text{ms}, taking account of the sampling frequency of 488~\text{kHz}. The data are transmitted to the personal computer via an Ethernet cable using socket communications. The FPGA board has two input channels. Ch.~1 acquires the interference signal from Photodetector 2, and Ch.~2 acquires the triangle wave signal applied to the mini fiber stretcher, which is appropriately attenuated. Fig.~\ref{fig11}(a) shows a plot of the 2-channel data acquired by the FPGA board.

\begin{figure}[tb]
\centering
\includegraphics[width=\linewidth]{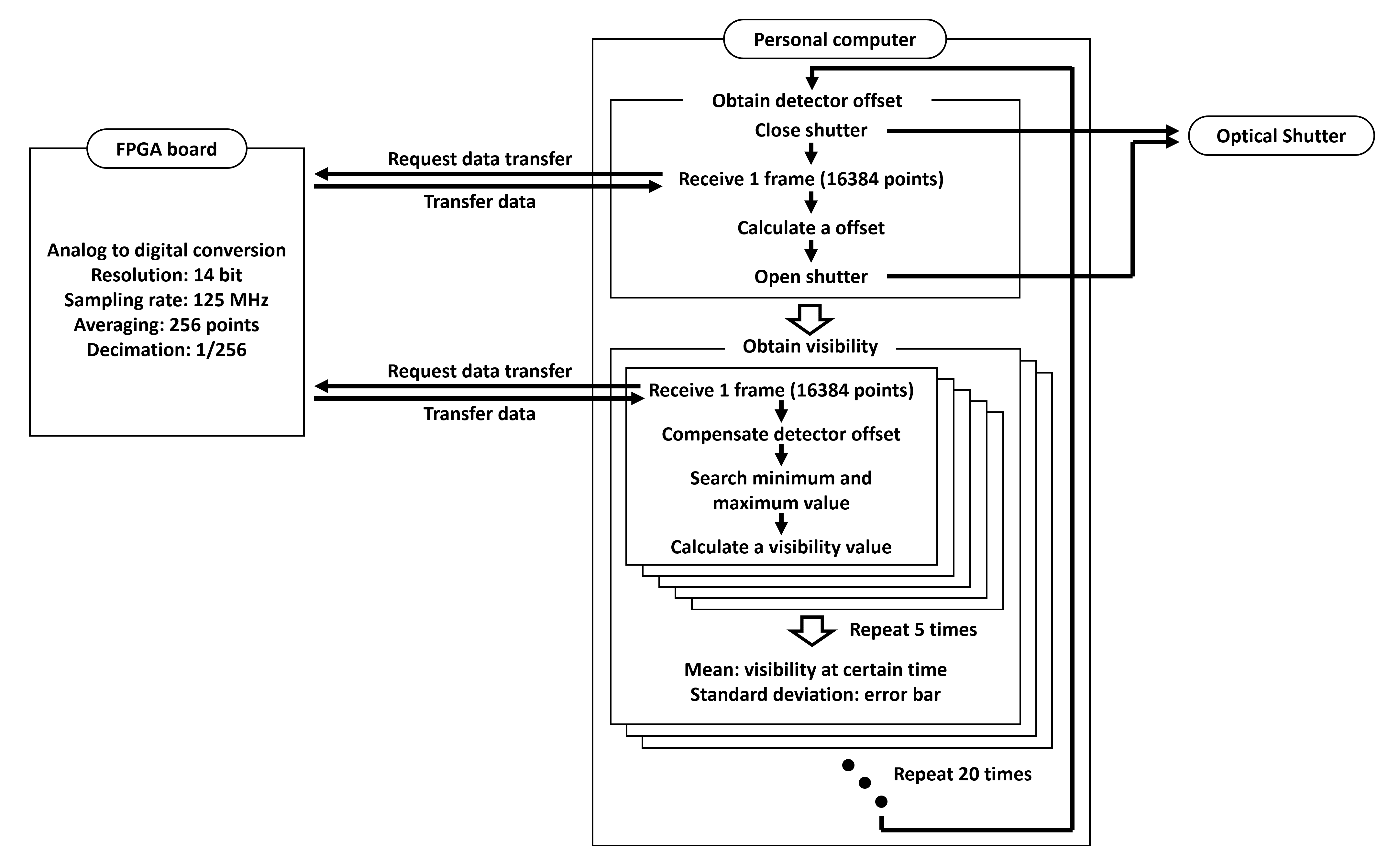}
\caption{Processes to obtain visibilities.}
\label{fig12}
\end{figure}

As shown in Fig.~\ref{fig12}, the visibility is calculated on the personal computer using the transferred data. First, the visibility value is calculated from each frame. Then, the visibility of one time is obtained by averaging the five visibility values from five frames, and its error bar is obtained by the standard deviation of them. Due to the time required for data transfer and real-time processing, a visibility value of one time is obtained about every 2.5 seconds.

Visibility $v$ is calculated from the maximum (max) and minimum (min) of the interference signal as
\begin{equation}
v=\frac{\rm{max}-\rm{min}}{\rm{max}+\rm{min}}. \label{eqn6}
\end{equation}
Here, max and min are the voltages that are measured from the voltage level without light.

The min and max of the interference signal are obtained from the area where the triangle wave signal is monotonically increasing (Fig.~\ref{fig11}(a)). Hence, the area is narrowed down to $\pm$3~\text{ms} around the time when the triangle wave signal crosses 0~\text{V}. Fig.~\ref{fig11}(b) and (c) show the min and max of the interference signal in Fig.~\ref{fig11}(a). In order to compensate for the slight DC level without light, the output voltage level without light is measured  by blocking Light 1 with a motor-driven optical shutter (gray points in Fig.~\ref{fig11}(b)). This voltage level without light is acquired in one frame and the average value is used as the reference voltage. The process to obtain voltage level without light is repeated every 20 time points of visibility measurements.

The visibility is calculated in real time, and the voltages applied to Fiber stretcher 1 and Fiber stretcher 2 are changed step by step to maximize the visibility. The visibility is maximized by trial and error as follows: First, Fiber stretcher 1 is moved in a certain direction by an appropriate step, and the visibility before and after the movement is compared. If the visibility improves, Fiber stretcher 1 is moved further, and this process is repeated until the visibility decreases. Next, the voltage applied to Fiber stretcher 2 is changed in the same way. Then, the driving target is back to Fiber stretcher 1, while the direction is opposite to the previous direction. By repeating these procedures, we maximize visibility. These processes are controlled by Python programs. In the experiment, we demonstrate not only the increase of visibilities by the trial-and-error processes but also the keeping of high visibilities by compensating for the drifts of polarizations by continuing the trial-and-error processes.

The step widths of the trial-and-error process explained above are set as follows: 30 mV when visibility is less than 99\%, 20 mV when visibility is between 99\% and 99.5\%, and 10 mV when visibility is greater than 99.5\%. Note that the FPGA board outputs DC voltages between $\pm$0.8~\text{V}. The $\pm$0.8~\text{V} outputs are amplified to about $\pm$3~\text{V} with op-amp ADA4898-1, and further amplified with a piezo actuator driver, adding about 400~\text{V} offset. Finally, the voltage applied to Fiber stretcher 1 ranges from 34~\text{V} to 772~\text{V}, and the voltage applied to Fiber stretcher 2 ranges from 110~\text{V} to 702~\text{V}.

\section{Experimental Result} \label{sec8}

\begin{figure}[tb]
\centering
\includegraphics[width=\linewidth]{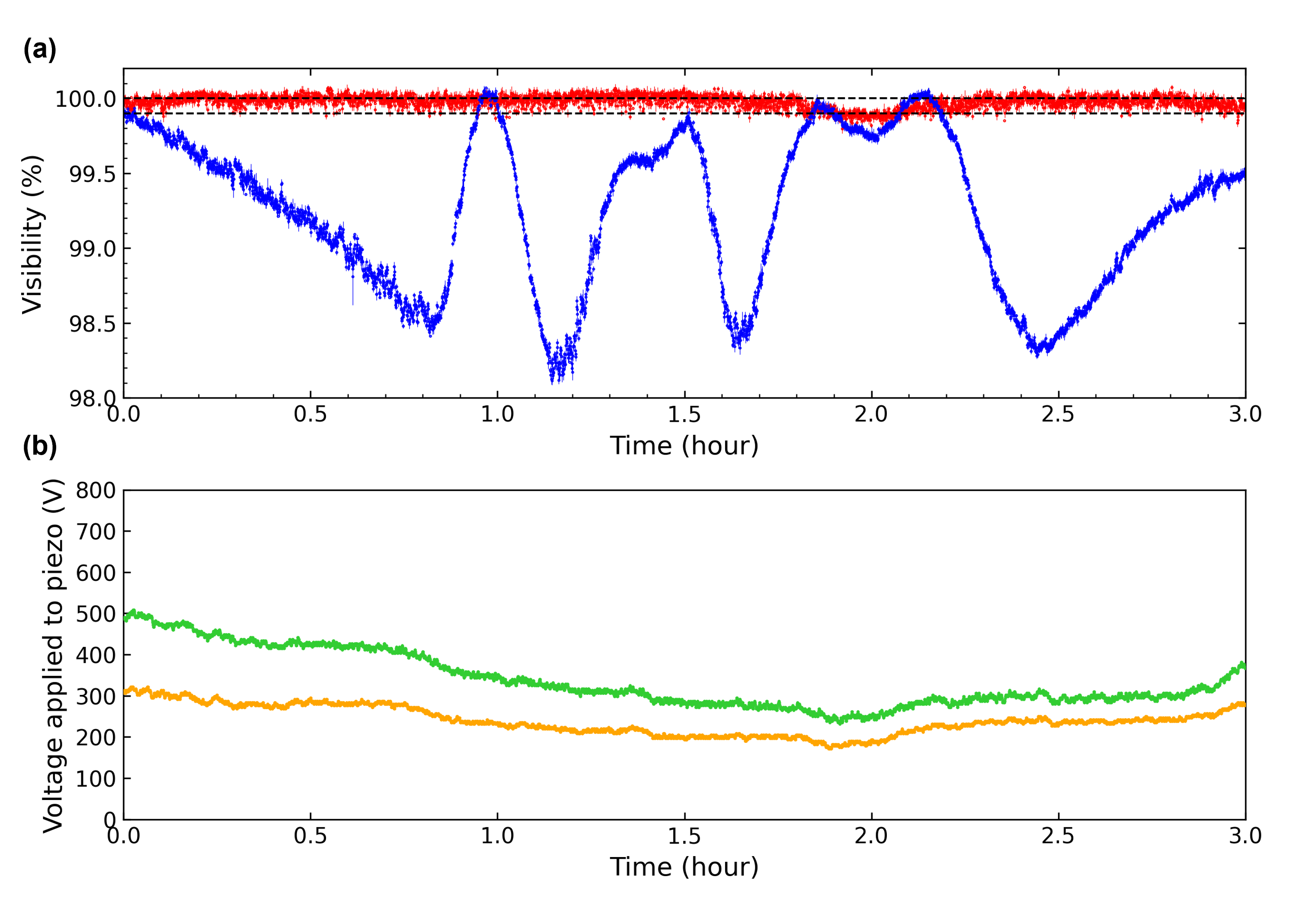}
\caption{(a) Time variation of visibility when visibility maximization by polarization control is turned on (red points) and turned off (blue points). Black dashed lines represent 100.0\% and 99.9\% visibilities. (b) Voltages applied to Fiber stretcher 1 (orange points) and Fiber stretcher 2 (green points) when visibility maximization is activated.}
\label{fig13}
\end{figure}

The visibility maximization experiment is performed continuously for three hours, and the results are as shown in Fig.~\ref{fig13}(a). The visibilities are red points for the case where visibility maximization by polarization control is activated and blue points for the case where it is deactivated for comparison. The horizontal axis is elapsed time and the vertical axis is visibility. Red points and blue points cannot be acquired at the same time. Hence, red points are firstly obtained for three hours, and then blue points for three hours. During the visibility maximization activated, visibilities are maintained basically above 99.9\%. On the other hand, during the visibility maximization deactivated, even though the initial visibility is set to almost 100\%, in three hours it drifts and temporarily drops to around 98\%. The error bars for each time are $\pm$0.02\% on average. These results show that the stability of the visibility is drastically improved by the CCC method.

A decrease in visibility can be evaluated as the angular mismatch of two linear polarizations. For the case the visibility maximization is activated, the visibility of 99.9\% corresponds to $0.9$ degrees of polarization mismatch. For the case the visibility maximization is deactivated, the minimum visibility of 98\% corresponds to $1.3$ Degrees of polarization mismatch. Note that the estimated worst case of visibility is 87\%, corresponding to the polarization mismatch of $\pm$10.3 Degrees, which are obtained by considering the finite PERs of used components. The experimental lowest visibility of 98\% is much above the estimated worst case of 87\%.

Fig.~\ref{fig13}(b) shows the voltages applied to Fiber stretcher 1 and Fiber stretcher 2, for the case where the visibility maximization is  activated. The orange points represent the voltage applied to Fiber stretcher 1, and the green points represent the voltage applied to Fiber stretcher 2. The voltages constantly change to maximize visibility. Since the voltage to draw the circular trajectory is 500~\text{V}, about 100~\text{V} change during the three-hour measurement shows that the polarization drifts by about one-fifth of the circular trajectories are compensated by the continuous trial-and-error processes.

\begin{figure}[tb]
\centering
\includegraphics[width=0.6\linewidth]{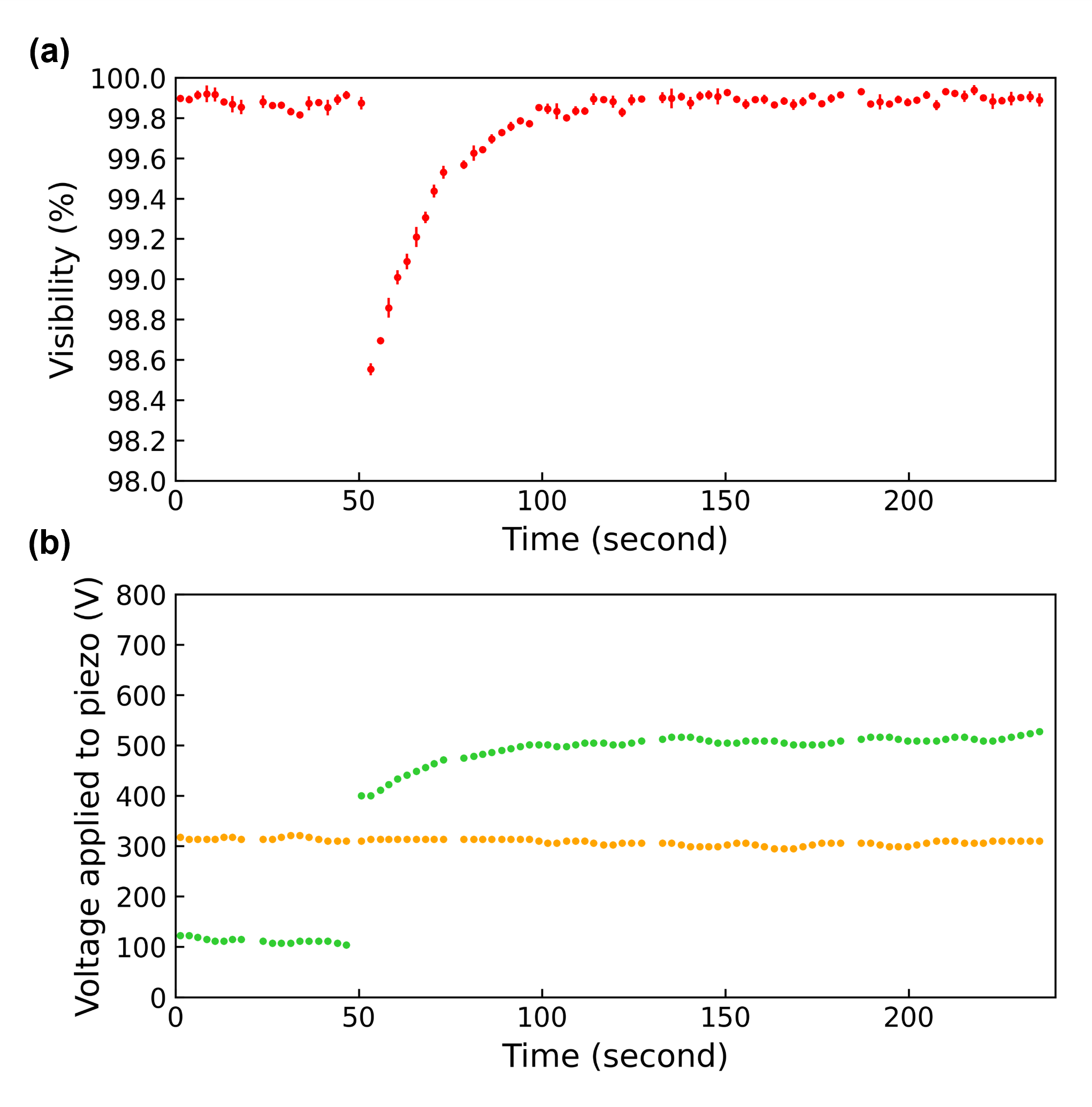}
\caption{Reset process of visibility maximization when the voltage applied to Fiber stretcher 2 reaches the lower limit. (a) Visibilities before and after the reset process. (b) Voltages applied to Fiber stretcher 1 (orange points) and Fiber stretcher 2 (green points). When the voltage applied to Fiber stretcher 2 reaches the lower limit, it is reset to 400~\text{V}.}
\label{fig14}
\end{figure}

The dynamic ranges of the fiber stretchers are finite. Thus, during long-term experiments to compensate for drifts of the polarizations, the voltages applied to the fiber stretchers may exceed the dynamic ranges, though such cases did not occur during the three-hour experiment in Fig.~\ref{fig13}. Our control program is written such that, when the voltage exceeds the dynamic range, the voltage is reset to the center of the dynamic range and then the visibilities are recovered by the trial-and-error processes. Fig.~\ref{fig14} shows the case where the voltage exceeds the dynamic range and the visibility is recovered. Fig.~\ref{fig14}(a) shows the visibilities, and Fig.~\ref{fig14}(b) shows the voltages applied to Fiber stretcher 1 (orange points) and Fiber stretcher 2 (green points). The voltage applied to Fiber stretcher 2 reaches the lower limit of 110~\text{V} at 50 seconds, and the voltage is reset to 400~\text{V}. Then the voltage gradually approaches an optimal point by the trial-and-error processes. The visibility decreases to about 98\% when the voltage is reset, and then gradually recovers to 99.9\%. It is also expected that this discontinuous visibility change can be rarer by substituting fiber stretchers with heaters with larger dynamic ranges.

We discuss the accuracy of the acquired visibilities. As is shown in Fig.~\ref{fig11}, the maximum  value of the interference signal is about 600 mV. From Eq.~(\ref{eqn6}), visibility decreases at a rate of 0.33\% when the minimum value increases by 1 mV. Since visibility is sensitive to the minimum value of the interference signal, it is important to obtain the minimum precisely. The FPGA board can acquire voltage signals with a range of ± 1~\text{V} and a resolution of 14 bits, where 1 bit corresponds to 0.12 mV. The circuit noise of the FPGA board is 0.10 $\text{mV}_\text{rms}$ when downsampled to 1/256. When Photodetector 2 is connected to the FPGA board, the circuit noise is 0.17 $\text{mV}_\text{rms}$. When light is incident and a DC voltage is 600 mV, the sum of optical shot noise and circuit noise is 0.18 $\text{mV}_\text{rms}$. Thus, the optical shot noise is almost negligible compared with the circuit noise. From the accuracy of the minimum values determined by circuit noises, it is assumed that the accuracy of the visibilities obtained from one frame data is about 0.06\%.

It is theoretically shown that fault-tolerant quantum computation is possible with more than 10~\text{dB} of squeezing levels\cite{Fukui2018}. A high squeezing level of 15~\text{dB} is reported where the visibility of homodyne measurements was 99.6\%\cite{Vahlbruch2016}. Thus, the visibility of 99.9\% with the CCC method is good enough to realize fault-tolerant quantum computation in the future.

\section{Comparison with conventional polarization controllers} \label{sec9}
There are various types of polarization controllers, but basically, they control polarization by applying stress to a non-polarization-maintaining fiber to induce birefringence. In this section, we compare the optical losses of the two types of manual fiber polarization controllers FPC032 (Thorlabs)\cite{ThorlabsFPC032}, HFPC-11-1300/1500-S-9/125-3A3A (OZ Optics)\cite{ozoptics_pol} and two types of motorized fiber polarization controllers MPC320 (Thorlabs), PCD-M02 (Luna)\cite{Luna}. The comparisons are summarized in Table \ref{tab1}.
\begin{table}[h]
 \centering
  \begin{tabular}{cccc}
   \hline
   Product Name & Fiber type & Manual/Electric & Loss \\
   \hline \hline
   FPC032 & SM & Manual & 0.4~\text{dB} (8\%) \\
   HFPC-11-1300/1500-S-9/125-3A3A & SM & Manual & 0.15~\text{dB} (3.4\%) \\
   MPC320 & SM & Electric & 3~\text{dB} (50\%)  \\
   PCD-M02 & SM & Electric & 0.08~\text{dB} (1.8\%) \\
   Homemade (used in this demonstration) & PM & Electric & 0.02~\text{dB} (0.5\%) \\
   \hline
  \end{tabular}
 \caption{Comparison of the performance of various polarization controllers.}
 \label{tab1}
\end{table}
FPC032 has three spools around which a non-polarization-maintaining fiber is wound several times to produce stress-induced birefringence. Then, by adjusting the angle of the spools, the direction of birefringence is changed to control polarization. This polarization controller requires about 5 m of non-polarization-maintaining fiber. If the fiber is wound tightly, the optical loss increases, but if the fiber is wound weakly, the birefringence becomes smaller, making polarization change difficult. The measured optical loss was 0.4~\text{dB} (8\%). In this product, microbends are considered to be generated at the base of the spool where the fiber is twisted, resulting in large optical losses. Instead of FPC032 where the angle of the spool is changed manually, MPC320 can be used where the angle of the spool can be changed electrically. However, MPC320 has a smaller spool radius, resulting in large optical losses. The actual optical loss of MPC320 was measured to be 3~\text{dB} (50\%).

HFPC-11-1300/1500-S-9/125-3A3A clamps a non-polarization-maintaining fiber to produce birefringence. The polarization is changed by the strength of the clamping force and the rotation angle of the clamp. The actual optical loss was measured to be 0.15~\text{dB} (3.4\%). Microbends are considered to be generated at the part where the single mode fiber is pressed and twisted by the clamp. In this product, the clamp is controlled manually.

PCD-M02 uses piezo actuators to press a non-polarization-maintaining fiber from various directions to produce birefringence. The non-polarization-maintaining fiber is pressed in a short distance of only a few millimeters, which causes microbends. The data sheet value of PCD-M02(Luna)\cite{Luna} polarization controller is less than 0.05~\text{dB} (1.1\%), and the measured value was 0.08~\text{dB} (1.8\%). This product controls polarization electrically.

In our homemade fiber stretcher used in the demonstration of the CCC method, birefringence is generated by stretching a 1.5 m fiber uniformly. Thus, microbends are considered to be less likely to occur and thus optical loss is reduced. In the demonstration, polarizations are controlled electrically.

\section{Conclusion} \label{sec10}
In this paper, we propose a method to maximize the interference visibility in a polarization-maintaining fiber system by controlling the polarizations to a crosspoint of the two circular trajectories on the Poincar\'{e} sphere, which we call the CCC method. Using fiber stretchers as polarization controllers, optical losses can be reduced. We also experimentally demonstrated the CCC method for three hours, and visibility was basically kept above 99.9\%. In contrast to the conventional methods where polarization is controlled by applying stresses to a non-polarization-maintaining fiber, the CCC method controls polarizations by pulling polarization-maintaining fibers with fiber stretchers, resulting in smaller microbends. Hence, the optical losses of the fiber stretchers were measured to be as low as 0.02~\text{dB} (0.5\%). Furthermore, in contrast to the conventional methods, we do not have to combine non-polarization-maintaining fibers with a polarization-maintaining fiber system. 

The CCC method has broad applications in situations where optical losses in fiber interferometer systems are undesirable. In particular, we can adapt this method to interference of loss-sensitive squeezed states, such as cluster states used for one-way quantum computation\cite{Menicucci2006, Yokoyama2013, Ukai2015, Larsen2019_2d, Larsen2021a}. Fiber systems do not require spatial alignment and are expected to improve stability over long periods of time. The CCC method can achieve polarization control electrically, which allows for automatic control. Furthermore, even if fiber systems have multiple interference points, visibilities can be optimized at all interference points by applying the CCC method from upstream to downstream of the systems.

\section*{Funding}
Japan Science and Technology Agency (JPMJMS2064, JPMJPR2254); Japan Society for the Promotion of
Science KAKENHI (18H05207, 20K15187).

\section*{Acknowledgments}
This work was partly supported by the UTokyo Foundation and donations from Nichia Corporation of Japan. T.N acknowledges financial support from Forefront Physics and Mathematics Program to Drive Transformation (FoPM).  M.E. acknowledges supports from Research Foundation for Opto-Science and Technology. 

\section*{Disclosures}
\noindent T.N and A.F. are inventors on a patent related to this work filed by the University of Tokyo on a priority date of 2 December 2021 with the Japan Patent Office (P). 


\bibliography{library}
\end{document}